\documentstyle[preprint,aps]{revtex}
\begin{document}
\title{{\bf
Deriving the quark condensate within a finite Fermi system from
the generating functional of 
chiral perturbation theory }}
\author{ M. Kirchbach\,$^1$ and A. Wirzba\,$^2$}
\address{ $^1$ Institut 
f\"ur Kernphysik, Universit\"at Mainz, D-55099 Mainz, Germany\\
$^2$ Institut f\"ur Kernphysik, TH Darmstadt,
 D-64289 Darmstadt, Germany}
\maketitle

\begin{abstract} 
\noindent 
The generating functional of heavy baryon
chiral perturbation theory at order ${\cal O}(Q^2)$ in the mean field
approximation (with a pseudoscalar source coupling which is consistent
with the PCAC-Ward identities on the current quark level) has been
exploited to derive Migdal's in--medium pion propagator.  It is shown
that the prediction for the density dependence of the quark condensate
obtained on the composite hadron level by embedding PCAC within the
framework of Migdal's approach to finite Fermi systems is identical to
that resulting from QCD.  
\end{abstract}
\vfill
{\small IKDA-96/7 \hfill (Published in Nucl. Phys. A 604 (1996) 395-405)}

\newpage

\section{\bf Introduction}

The evolution of the quark condensate,
$\langle\,  0 \mid\bar q q\mid 0 \,\rangle$,
with density, $\rho $,
is one of the most intriguing and controversial problems
of intermediate hadron physics that had attracted much attention
during the last years partly because of
its relevance for chiral symmetry restoration at finite densities.
On the one hand, QCD--inspired quark models predict an in fact
model independent linear decrease of the quark condensate
in accordance with
$\langle \widetilde{0}\mid \bar q q\mid \widetilde{0}\rangle 
= \langle 0\mid \bar q q\mid 0\rangle (1 -\sigma_N\rho /f_\pi^2 m_\pi^2)$
\cite{condensate,veron,Cohen,lutz}. 
Here $m_\pi$ and $f_\pi$  in turn denote the mass and weak decay constant
of the pion, whereas $\sigma_N$ stands for 
the pion--nucleon sigma term.
In the following ``tilde'' will be used to denote in--medium states.
On the other hand, attempts
have been done to deduce the properties of quark matter
at finite densities from the properties of
composite hadrons at finite densities \cite{DEE 92,CEK 94}
with the aim to extend the linear decrease of $\langle \,\widetilde{0}\mid
\bar q q\mid\, \widetilde{0}\rangle $ with $\rho $ (which was mentioned above)
to higher powers of the matter density.
The scheme exploited is based on the evaluation of 
the Gell--Mann--Oakes--Renner (GOR) relation 
(see \cite{Birse} for a recent review).
The latter 
relates the quark condensate to the divergence of the charged axial
vector current and can be evaluated
in exploiting PCAC in combination with Migdal's pion propagator.
The basic ingredient of such schemes was
the assumption on the linear dependence of the S--wave pion self--energy
on matter density.
We here show that if 
Migdal's propagator is derived
from the generating functional of heavy baryon
chiral perturbation theory at order ${\cal O}(Q^2)$ in the mean field
approximation, 
the S--wave pion self--energy is no longer linear in $\rho $ but contains 
an infinite number of higher powers of matter density. With that
the prediction for the density dependence of the quark condensate
obtained on the composite hadron level by embedding PCAC within the
framework of Migdal's approach to dense matter is shown to be
identical to that resulting from QCD.
In this way, the possibility to learn more on the evolution of 
the quark condensate
with density from the in--medium composite hadron physics is ruled out.
Through the  
pseudoscalar source coupling used by us consistency is ensured
between PCAC on the composite hadron level, on the one side,
and PCAC-Ward identities on the current quark level, on the other side.
 
The paper is organized as follows. In the next section we review shortly
the basic ingredients of chiral perturbation theory
in the heavy baryon limit and in the mean field approximation,
present the corresponding effective chiral lagrangian, and derive
Migdal's pion propagator.
In sect.~3 we evaluate the GOR relation within a finite Fermi system
before closing with a short summary and discussion.

\section{Deriving Migdal's pion propagator from
a chiral effective lagrangian}

\noindent Chiral perturbation theory \cite{GL 84} was developed as a
tool for constructing Green functions in QCD. It is based on the
assumption that the fundamental symmetry of QCD, the chiral symmetry,
is realized in the Nambu--Goldstone mode with the pions acting as
Goldstone bosons.  As long as Goldstone particles interact only weakly
with each other and the matter (fermion) fields, it is possible to
expand correlation functions in powers of the light quark masses and
the external pion momenta thought to be very small on the hadronic
scale of $\Lambda\approx 1GeV$.  An $SU(3)_L\otimes SU(3)_R$ chiral
effective lagrangian (subsequently denoted by ${\cal L}_{\rm MG}$)
containing the characteristic momenta to second order (so called
``next--to--leading'' order) was constructed by Manohar and Georgi
\cite{MG 84} in the heavy-baryon formalism or static approximation.
In the $SU(2)_L\otimes SU(2)_R$ reduction and in the $S$--channel the
latter lagrangian can be expanded in the pion field and cast into the
following form \cite{BKM 93,Br 93} which is valid to order ${\cal
O}(Q^2)$ with $Q$ standing for the external pion momentum or mass:
\begin{eqnarray} 
 {\cal L}_{\rm MG} &=& i\bar N(x) (v\cdot
  \partial)N(x)
                   -\sigma \bar N (x)N(x)
         +{1\over 2}\left(\partial\, \pi (x)\right)^2 
         -{1\over 2}m_\pi^2\,\pi(x)^2 
                \nonumber\\
                  & +&{1\over f_\pi^2}\left({1\over 2}\sigma \pi(x)^2
                   +c_2(v\cdot\partial\pi (x))^2
                  + c_3  (\partial \pi (x))^2\right)\,\bar N(x)N(x) 
 + \cdots \, .
    \label{L_MG}
\end{eqnarray}
The quantity $v_\mu$ stands for the four-velocity of the nucleon ($N$)
in the heavy baryon limit, and reduces to $v_\mu = (1,0,0, 0)$ for a
nucleon at rest, whereas $\pi^a(x) $ stands for the pion field 
at zero matter density.
The constant coefficient $\sigma $ is linear in the quark masses and
therefore of the order ${\cal O}(Q^2)$.  It serves to {\em increase\/}
the nucleon mass over its $SU(2)$ chiral limit value of $m_0= 890 MeV$
to $m_N=m_0+\sigma > m_0 $.  By this requirement the sign of $\sigma $
is fixed to be positive.  Thus modulo ${\cal O}(Q^3)$ corrections, the
identification $\sigma = \sigma_N$, where $\sigma_N$ is the
pion--nucleon--sigma term, is natural~\cite{GLS 91}.  The combination
$(c_2 + c_3)m_\pi^2$ can then be extracted (on the tree level) from
the empirical isospin even pion nucleon scattering length $a_{\pi
N}^{+}$ as $(c_2 + c_3)m_\pi^ 2 \approx -26\,{\rm MeV}$~\cite{BKM 93}.
Note, the coefficients $c_2$ and $c_3$ differ from those defined in
ref.\,\cite{BKM 93} by ${\cal O}(Q^3)$ terms.  The relevant generating
functional at the order ${\cal O}(Q^2)$ reads 
\cite{GL 84,GSS 88,ThWi 95} 
\begin{equation}  
e^{i Z_{\rm eff}\lbrack s,
                     p, V_\mu ,A_\mu,\eta ,
 \bar \eta \,\rbrack}
 = {\cal N}\int dUdNd\bar N \, \exp \left(i\int d^4x\, \left \{ 
   {\cal L}_{\pi \pi}\mbox{+}{\cal L}_{\pi N}\mbox{+} 
 \bar \eta N\mbox{+}\bar N \eta\right \}\right)\, . \label{Zeff}
\end{equation}
Here, ${\cal N}$ is an overall normalization factor, $U, N, {\bar N}$
are dummy integration fields for the pions in the non-linear
representation and the nucleons in the heavy baryon formulation,
respectively. The lagrangians ${\cal L}_{\pi \pi}$ and ${\cal L}_{\pi
N}$ are defined in such a way that they do not only depend on the
fields, but also on the scalar ($s$), pseudoscalar ($p$), vector
($V_\mu$) and axial vector ($A_\mu$) sources.  The sources $\bar \eta
$ and $\eta $ generate one--nucleon in-- and out--states,
respectively.  The lagrangian entering the generating functional
(\ref{Zeff}), ${\cal L}_{\rm GSS}:= {\cal L}_{\pi \pi}+{\cal L}_{\pi
N}$ (which we denote the lagrangian of Gasser, Sainio and
\v{S}varc~\cite{GSS 88}), is given by the nucleon kinetic energy term
(which is of leading order, ${\cal O}(Q)$) and to subleading order,
${\cal O}(Q^2)$, by~\cite{BKM 93,ThWi 95}
\begin{eqnarray}
 {\cal L}^{(2)}_{\pi \pi}&=& \frac{f_\pi^2}{4} 
   {\rm Tr}\left( (\nabla^\mu U)^\dagger \nabla_\mu U \right )
 +\frac{f_\pi^2}{4} {\rm Tr}( U^\dagger \chi\mbox{+}\chi^\dagger U)\, , 
  \nonumber \\
 {\cal L}^{(2)}_{\pi N}&=& 
 c_2 {\bar N}\! 
 \left(iu^\dagger \nabla_\mu U u^\dagger\, v^\mu\right)^2 \!\! N
 +c_3 {\bar N}\! \left(i u^\dagger\nabla_\mu U u^\dagger \right)^2 
  \!\! N
 -\frac{\sigma}{4 m_\pi^2} \bar N N\,
 {\rm Tr}(U^\dagger \chi\mbox{+}\chi^\dagger U) \, 
 , \nonumber \\
 \nabla_\mu U &:=&\partial_\mu U - i{1\over 2}\left\{ \tau^a A^a_\mu , 
  U\right\}
             -i{1\over 2}\left [\tau^a V^a_\mu , U\right] \, ,\nonumber \\
    \chi&:=&2B(s+i\tau^a p^a) = 2 \frac{ m_\pi^2}{m_u +m_d} 
                               (s+i\tau^a p^a) \, ,\nonumber
\end{eqnarray} 
where $U$=$u^2$=$\exp(i\tau^a\pi^a/f_\pi)$ and $m_u$ and $m_d$ are the
up-- and down-quark masses, respectively. Note that the sources are
coupled in such a way that the lagrangian (including the sources) is in fact even {\em locally} chiral invariant \cite{GL 84}.
Setting the vector and axial vector sources, $V_\mu$ and $A_\mu$, to
zero, using the scalar source $s$ to generate the quark mass matrix
$s$=${\cal M}$=${\rm diag}(m_u,m_d)$ and finally expanding $U$ to
second order in $\pi^a$, one gets the following relation between the
${\cal L}_{\rm GSS}$ lagrangian and ${\cal L}_{\rm MG}$ of (\ref{L_MG}):
\begin{equation}
{\cal L}_{GSS}|_{A_\mu=V_\mu=0,s={\cal M}}=
 {\cal L}_{\rm MG} +
 \left(1-{{\sigma_N \bar N(x)N(x)}\over 
 {f_\pi^2m_\pi^2}}\right)j^a(x)\pi^a (x)
 \; .
\end{equation}
Here, we introduce a ``renormalized'' pseudoscalar isovector $j^a$
which is related to the original source $p^a$ by
$j^a(x) = 2Bf_\pi\, p^a(x) =g_\pi p^a(x)$ 
where $g_\pi$ is the vacuum coupling constant of the pseudoscalar
density to the pion~\cite{GL 84}, $\langle\, 0 \mid \bar q i \gamma_5
\tau^a q \mid
 \pi^b \, \rangle = \delta^{ab} g_\pi$.  Note that via the generating
functional formalism of \cite{GL 84} (where the symmetry-breaking
quark--mass terms are accounted for by the scalar source $s$) the
chirally symmetric structure of the source couplings at the QCD level
(which is even {\em local} if the axial vector and vector source
transform covariantly) is manifestly kept at the effective lagrangian
level -- in case all sources are non--zero, of course. Thus the
pseudoscalar source $p^a$ at the effective lagrangian level is to be
identified at the QCD level with the source which directly couples to
the pseudoscalar quark density. Therefore the PCAC-Ward identities
between the amplitudes involving on the one hand the axial current and
on the other hand the pseudoscalar density are kept unaltered in the
framework of the effective lagrangian ${\cal L}_{\rm GSS}$. This
lagrangian will automatically generate the PCAC-consistent amplitudes,
e.g.\ the isospin-symmetric $S$--wave $\pi N$ amplitudes at the
Weinberg, Adler, Cheng-Dashen point and at threshold, see
ref.\,\cite{ThWi 95}. All the measurable quantities, however, are of
course independent of the special choice of the pseudoscalar source
coupling (see ref.\,\cite{CWZ 69}) 
and will follow from ${\cal L}_{\rm GSS}$ as well as
${\cal L}_{\rm MG}$ or any other variant based on ${\cal L}_{\rm
MG}$~\cite{ThWi 95}.

Now in the mean field limit, where $ \bar N(x) N(x)$ (which is equal
to $N(x)^\dagger N(x)$ in the heavy-baryon formulation) is
approximated by the local matter density $ \rho (x)$, the chiral
lagrangian ${\cal L}_{\rm GSS}$ if rewritten in terms of the $U$
field takes the manifestly chirally symmetric form (see \cite{ThWi 95},
\cite{WiTh 95} for details) :
\begin{equation}
 {\cal L}_{\rm MF} = { {f_\pi^2 }\over 4}
 (g^{\mu\nu }\mbox{+}{{D^{\mu\nu }\rho}\over f_\pi^2})
 \, {\rm Tr} \left( (\nabla_\mu U)^\dagger \nabla_\nu U \right)
  +{f_\pi^2\over 4}(1\mbox{$-$}{{\sigma_N \rho}\over {f_\pi^2m_\pi^2}})\,
 {\rm Tr}(U^\dagger\chi\mbox{+}\chi^\dagger U) +\cdots \, , 
 \label{L_MF}
\end{equation}
with $D^{\mu \nu } := 2c_2v^\mu v^\nu +2c_3g^{\mu\nu}$. 
The weak axial vector decay constant of the in-medium pion is the
time--like one because of the special case of the $S$--channel
considered and is identified from the first term in eq.\,(\ref{L_MF})
as
\begin{equation}
 \left(f_\pi^t(\rho)\right)^2 
 = f_\pi^2 \left(1+ {{2(c_2+c_3)\rho }\over 
  {f_\pi^2}}\right)\, 
 + {\cal O} (m_\pi )\, .  \label{f_pi_t}
\end{equation}
More precisely, it is extracted from the residuum of time-like axial 
current--current two-point function
\begin{eqnarray} 
 \frac{\delta}{\delta A^a_0(-q)} \frac{\delta}{\delta A^b_0(q)} \, 
  Z_{\rm MF}|_{A_\mu=V_\mu=p=0,s={\cal M}} \nonumber
\end{eqnarray}
where $Z_{\rm MF}$ is the to $Z_{\rm eff}$ analogous generating
functional in the mean-field approximation --- see ref.\cite{ThWi 95}
for more details.  In complete analogy, the in--medium condensate can
be read off from the second term of eq.\,(\ref{L_MF}) as follows:
\begin{equation}
 \langle\, \,\widetilde{0}\mid \bar q q \mid \widetilde{0}\,\rangle
  =\langle\,\, 0\mid \bar q q\mid 0\,\rangle 
  \left (1-{{\sigma_N\rho}\over {f_\pi^2m_\pi^2}}\right)
  + {\cal O}( m_\pi)\, .
     \label{cond_med} 
\end{equation}
Here, the
vacuum quark condensate is given by
$ \langle\, 0\mid \bar q q \mid 0 \,\rangle  =-2f_\pi^2 B
+ {\cal O}(m_\pi^2)$. 
Again, the precise derivation is based on the generating functional 
techniques:
\begin{eqnarray}
     \frac{\delta}{\delta s(x)} Z_{\rm MF}|_{
 A_\mu{=}V_\mu{=}p{=}s{=}0}
 =- \langle\, \widetilde{0} 
      \mid \bar u u + \bar d d \mid \widetilde{0}\,\rangle 
 \nonumber
\end{eqnarray}
(see ref.\cite{ThWi 95}).  The propagator of a charged pion in the
medium is developed from the mean-field analog of the generating
functional (\ref{Zeff}) in the standard way as
\begin{eqnarray}
 \frac{\delta}{\delta j^a(q)} \frac{\delta}{\delta j^b(-q)} 
  Z_{\rm MF}
 |_{A_\mu{=}V_\mu{=}j{=}0,s{=}{\cal M}} =
 \Delta_{\Phi_\pi}(q^2,\rho)\, \delta_{ab} \, ,  \nonumber
\end{eqnarray}
with
\begin{eqnarray} 
 \Delta_{\Phi_\pi}(q^2,\rho)
  &=&{ {(1- {{\sigma_N\rho}\over {f_\pi^2m_\pi^2}})^2}
 \over {q_0^2(1 + {{2(c_2+c_3)\rho}\over f_\pi^2})
 -{\vec{q}\,}^2 (1 - {{2c_3\rho}\over f_\pi^2})
 -m_\pi^2 (1 - {{\sigma_N\rho}\over {f_\pi^2m_\pi^2}})}}
 + {\cal O}(m_\pi^3)\, \label{GSS_prop}\\ 
 &=& \frac{\left( g_\pi^\ast(\rho)/g_\pi \right)^2}
       { q_0^2 -{{\vec q}\,}^2 \frac{1-2c_3\rho/f_\pi^2}
 {1+2(c_2+c_3)\rho/f_\pi^2}- {m_\pi^\ast}^2} + {\cal O}(m_\pi^3) \, ,
\end{eqnarray}
where $g_\pi^\ast(\rho)$ is the in-medium coupling constant of the
pseudoscalar density to the pion~\cite{ThWi 95}
\begin{eqnarray}
  \left ( g^\ast_\pi(\rho) \right)^2= g_\pi^2 
   \frac{ \left( 1 -\frac{\sigma_N\rho} {f_\pi^2 m_\pi^2}\right)^2}
    { 1 +\frac{2(c_2+c_3) \rho}{f_\pi^2}}\, 
 . \label{g_pi}
\end{eqnarray}
The in--medium pion mass $m^\ast_\pi$ is determined by the
pole of this propagator and is given by (see refs.\cite{DEE 92,ThWi 95})
\begin{equation}
 \left(m_\pi^\ast\right)^2 = m_\pi^2 \,
 \frac{ 1- \frac{\sigma_N\rho }{f_\pi^2m_\pi^2}} 
 {1 + \frac{2(c_2+c_3)\rho }{f_\pi^2} }
 \approx m_\pi^2 \, . \label{m_pi_star}
\end{equation} 
{}From eqs.\,(\ref{f_pi_t}--\ref{cond_med}) (~with 
$m_\pi^2=B(m_u\mbox{+}m_d)\mbox{+}{\cal O}(m_\pi^4)$~) 
and eq.\,(\ref{m_pi_star}) 
the explicit in--medium extension of the
Gell--Mann--Oakes--Renner relation immediately follows as
\begin{eqnarray}
 \left(f_\pi^t(\rho )\right)^2\, \left(m_\pi^\ast\right)^2 
 & =& -2m_q \langle\, \widetilde{0}\mid 
  \bar q q\mid \widetilde{0}\,\rangle + {\cal O}(m_\pi^3)\nonumber\\
 & = & f_\pi^2m_\pi^2 \left(1 -{{\sigma_N\rho }\over 
 {f_\pi^2m_\pi^2}}\right)
 + {\cal O}(m_\pi^3) \, , \label{GOR_rel}
\end{eqnarray}
where $m_q$ stands for the averaged quark mass~\cite{ThWi 95}.  
Finally, the PCAC-consistent pion field at finite density is 
related to the pion field at zero density according to
\begin{equation}
 \widetilde{\pi}^a(\rho )= 
   {{\delta Z_{\rm MF}\lbrack j^a, \rho  \rbrack}\over {\delta j^a}} \, 
   = (1-{{\sigma_N\rho}\over {f_\pi^2m_\pi^2}})\pi^a \, .
 \label{pi_class}
\end{equation} 
Whereas the denominator of the propagator (\ref{GSS_prop}) and
especially its pole position is independent of the GSS-choice for the
pseudoscalar source coupling and therefore also of PCAC, the numerator
and $g_\pi^\ast(\rho)$ result from the specific structure of the
pseudoscalar source coupling and thus are scheme--dependent, here
PCAC--scheme--dependent.  Note that in case the effective lagrangian
${\cal L}_{\rm GSS}$ were replaced by the lagrangian ${\cal L}_{\rm
MG}$ plus a trivial nucleon-independent coupling to the pseudoscalar
source, $j^a(x)\pi^a(x)$, the term $(1-\sigma_N \rho/f_\pi^2
m_\pi^2)^2$ in the numerator of the propagator (\ref{GSS_prop}) and in
the definition of $g_\pi^\ast(\rho)$ in eq.\,(\ref{g_pi}) 
would be
absent, indicating that the ``MG'' choice for the in--medium pion field,
$\pi_{\rm MG}^a(\rho) = \pi^a$,
is not respecting PCAC.  The results of
eqs.(\ref{f_pi_t}--\ref{cond_med}) and
(\ref{m_pi_star}--\ref{GOR_rel}) which all refer to measurable
quantities do of course not depend on the PCAC choice for the
in-medium pion field (\ref{pi_class}) (as the pseudoscalar source
$p^a$ or $j^a$ did not enter in any of the derivations) and would
follow from the ``MG''-lagrangian as well~\cite{ThWi 95}.

Now one can compare the expression (\ref{GSS_prop}) with the standard form of
Migdal's propagator in the finite Fermi systems theory (FFS) 
\cite{Achm 80}
\begin{equation}
 (\Delta_\pi)_{\rm FFS} (q^2,\rho) =
 \left \{ q_0^2 \left(1-{{\partial\Pi^S}\over {\partial q_0^2}}\right)
 -\gamma (\rho)  \vec{q}\, ^2 -
 m_\pi^ 2 \left(1 +{{\Pi^S(0,0)}\over m_\pi^2}   \right)\right\}^{-1}\, ,
 \label{Mig}
\end{equation}
where the $S$--wave pion self energy $\Pi^S(q_0,\vec q\mbox{=}0)$ has been
expanded around $q_0^2=0$ as $\Pi^S(q_0,0)=\Pi^S(0,0)
+\frac{\partial\Pi^S}{\partial q_0^2} q_0^2$.
By doing so, 
one can immediately read off the following relations consistent with 
the PCAC--Ward identities:
\begin{eqnarray}
 \gamma (\rho ) & = &
 {{1- {{2c_3\rho }\over f_\pi^2} }\over
 {(1 -{ {\sigma_N\rho }\over {f_\pi^2m_\pi^2}})^2 } }\, ,
 \label{Mig_gam}\\
 1-{ {\partial\Pi^S}\over {\partial q^2_0}}
 &=&
 {{1+ {{2(c_2+c_3)\rho}\over f_\pi^2}}\over
 {(1 -{ {\sigma_N\rho }\over {f_\pi^2m_\pi^2}})^2 } }
 \approx {1\over {1-{{\sigma_N\rho}\over {f_\pi^2m_\pi^2}}}}\, ,
 \label{Mig_dPi}
 \\
 1+{{\Pi^S(0,0)}\over m_\pi^2} &=&
 {1\over {1 -{{\sigma_N\rho }\over {f_\pi^2m_\pi^2}}}}\, ,
 \label{Mig_Pi}
\end{eqnarray}
where in eq.\,(\ref{Mig_dPi}) the following approximate
relation has been used: $ 2(c_2 +c_3) m_\pi^2 \approx -\sigma_N$.
Note that $ (1-{ {\partial\Pi^S}\over {\partial q^2_0}})^{-1}
=(g_\pi^\ast(\rho)/g_\pi)^2$ as expected.
In approximating 
\begin{equation}
 \left(1-{{\sigma_N\rho}\over {f_\pi^2m_\pi^2}} \right)^{-1}
 \approx 1+{{\sigma_N\rho}\over {f_\pi^2m_\pi^2}} \, ,
\end{equation}
and inserting into eqs.\,(\ref{Mig_dPi}-\ref{Mig_Pi}) 
the standard (original) parametrizations are recovered:
\begin{equation}
 \Pi^S(0,0)\approx {{\sigma_N\rho }\over f_\pi^2 }\approx
 -{{\partial \Pi^S}\over {\partial q_0^2}}\, . \label{Pi_appr}
\end{equation}
The relations (\ref{Mig_gam}--\ref{Mig_Pi}) show that 
Migdal's pion field has to be identified with 
$\widetilde{\pi } (\rho )$ from eq.\,(\ref{pi_class}).

\section{\bf The quark condensate within a finite Fermi 
system}

{\em One possibility\/} for evaluating the GOR relation on the
composite hadron level is to assume validity of {\em PCAC as an
operator relation\/}, $\partial ^\mu
 \widehat{A}_\mu^a = f_\pi m_\pi^2 \widehat{\pi}^a$,
and to exploit the Migdal propagator (under the identifications
(\ref{Mig_gam}--\ref{Mig_Pi}) now shown to be consistent with PCAC)
for the in--medium pion. In doing so, the following relation is
obtained \cite{Birse}:
\begin{eqnarray}
 \Pi^{\rm GOR} =\lefteqn{ {\lim }_{q_0\to 0, \vec{q}\to 0}\ 
 \frac{i}{3}\int d^4x\, e^{-iq\cdot x}\langle\, \widetilde{0}\mid
 T\partial^\mu \widehat{A}_\mu^a(x) \,
   \partial^\nu \widehat{A}_\nu^a(0)
 \mid\widetilde{0}\,\rangle \nonumber} \\
 &=&{\lim }_{q_0\to 0,\vec{q}\to 0}\ \frac{i}{3} f_\pi^2m_\pi^4 
 \int d^4 x \,e^{-iq\cdot x}
 \langle\, \widetilde{0}\mid T \widehat{\pi}^a(x) \,  
 \widehat{\pi}^a(0) \mid \widetilde{0}\,\rangle 
\nonumber\\
 &= & {\lim }_{q_0\to 0 }\,
 {{f_\pi^2m_\pi^4}  \over
 { \left(1 - {{\partial\Pi^S}\over {\partial q_0^2} }\right)q^2_0 
 - m_\pi^2 \left( 1 +{{\Pi^S(0,0)}\over m_\pi^2}\right) } }
 = -\frac{f_\pi^2m_\pi^2}{1 + \frac{\Pi^S(0,0)}{m_\pi^2}}\, .
 \label{GOR_op}
\end{eqnarray}
On the other side, in the $\vec{q}\to 0$ limit PCAC is determined only
by the partial time derivative of the axial charge density which can
be replaced by 
\begin{equation}
\partial^\mu \widehat{A}_\mu^a
 \stackrel{ {\vec{q}\to 0}}{\longrightarrow} \partial^0 \widehat{A}_0^a
 =i \lbrack {\cal H}, \widehat{A}_0^a \rbrack\, , 
\end{equation}
where
${\cal H} $ stands for the QCD hamiltonian density. By integrating by
parts one can then cast the GOR correlator into the form
\begin{equation}
 \Pi^{\rm GOR}={1\over 3}\Sigma_{a=1}^3 \langle\, \widetilde{0}\mid
 \lbrack \widehat{A}_0^a,\lbrack \widehat{A}_0^a,
 {\cal H}\rbrack\rbrack\mid 
 \widetilde{0} \,\rangle\, .
\end{equation}
As long as the chiral symmetry violating term in the QCD lagrangian
corresponds to the non--zero current quark masses, $\Pi^{\rm GOR}$ can
be expressed in terms of the averaged current quark mass $m_q$ and the
quark condensate at non--zero density $\langle\,\widetilde{0} \mid
\bar q q\mid \widetilde{0}\,\rangle$ according to
$\Pi^{\rm GOR} =2m_q\langle\, \widetilde{0}\mid \bar q q\mid 
\widetilde{0}\,\rangle$,
where the flavor independence of the condensate has been assumed.
Together with eqs.\,(\ref{GOR_op}) and the PCAC-consistent
identification (\ref{Mig_Pi}) the last equation leads to the following
results
\begin{eqnarray}
 2m_q\langle\, \widetilde{0}\mid \bar q q\mid \widetilde{0}
   \,\rangle  &= &
 - \frac{f_\pi^2m_\pi^2}{1+ \frac{\Pi^S(0,0)}{m_\pi^2}} 
 \label{cond_exa}
\\
 & =& -\left (1 - \frac{\sigma_N\rho}{f_\pi^2m_\pi^2}\right)
  f_\pi^2m_\pi^2 
 =  -\left(f_\pi^t(\rho)\right)^2 \,\left(m_\pi^\ast\right)^2
   \, .
 \label{cond_lin}
\end{eqnarray}
In this way (~with the help of the PCAC-consistent relation
(\ref{Mig_Pi})~) eq.\,(\ref{GOR_rel}) is exactly reproduced -- in the
PCAC scheme on the operator level.  If the approximations from
eq.\,(\ref{Pi_appr}) were inserted for $\Pi^S(0,0)$ into
eq.\,(\ref{cond_exa}) a seemingly hyperbolic decrease of the quark
condensate with density would be obtained in line with the result
reported in~\cite{CEK 94}.  The reason for this incorrect
interpretation of eq.\,(\ref{cond_exa}) is an inconsistent treatment
of PCAC on the quark level.  The assumption of the validity of PCAC on
the operator level for composite hadrons should not be combined with
the approximations of eq.(\ref{Pi_appr}) which do destroy the form of
the Migdal propagator as developed in consistency with PCAC on the
current quark level.  Note that there is no compelling reason to
derive the GOR relation by exploiting the PCAC hypothesis.  If,
however, a definite scheme as e.g.\ PCAC has once been chosen, then
all the quantities have to be calculated only in that very scheme.

A {\em second possibility\/}, to derive the density dependence of the
quark condensate within Migdal's theory of the finite
Fermi systems is to calculate the PCAC condition
on the {\em matrix element level\/} rather than postulating
its validity as operator relation.  This is much closer than the first
approach to the generating functional approach which correlates the
{\em vacuum expectation values} of time-ordered products of
field-operators.  The effect of the polarization of the medium during
the propagation of a $S$-- / $P$--wave pion on the respective pion
weak decay constants (in turn denoted by $( f_\pi^{\,S})_{\rm FFS}$
and $(f_\pi^{\,P})_{\rm FFS}$) is expressed by means of the following
parametrizations of the matrix elements of the pion weak axial vector
 current $\widehat{A}^a_{\mu (\pi )}$:
\begin{eqnarray}
 \langle\,\widetilde{0} \mid \vec{\widehat{A}^a}_{\!\!(\pi)} 
 \mid \widetilde{\pi}^b \,\rangle
 &=:& (f_\pi^P)_{\rm FFS} (\rho )\  i\vec{q\, }\,\delta^{a b} \, , 
  \label{A_pi_vec}\\
 \langle\,\widetilde{0} \mid 
 \widehat{A}^a_{0(\pi )}\mid\widetilde{\pi}^b \,\rangle
  &=:& (f_\pi^S)_{\rm FFS} (\rho )\ iq_0\, \delta^{a b}\, , 
    \label{A_pi_0}\\
\langle\,\widetilde{N}(\vec{p}_2)\mid 
 \vec{\widehat{A}^a}_{\!\!(\pi)} 
\mid \widetilde{N}
 (\vec{p}_1)\,\rangle &=:& i\vec{q}\,(f^P_\pi)_{\rm FFS}(\rho )\,
 \langle\, \widetilde{N}(\vec{p}_2)\mid
  \widehat{\pi}^a\mid \widetilde{N}(\vec{p}_1)\,\rangle\,
  ,\\
\langle\, \widetilde{N}(\vec{p}_2)\mid \widehat{A}^a_{0(\pi)}\mid
  \widetilde{N}(\vec{p}_1)\,\rangle  
  &=:&
 iq_0\, (f^S_\pi )_{\rm FFS}(\rho )\,
  \langle\,\widetilde{N}(\vec{p}_2)\mid
 \widehat{\pi}^a\mid \widetilde{N}(\vec{p}_1)\,\rangle\, ,
 \label{A_pi_0NN}
\end{eqnarray}
where $q^\mu = p_1^\mu - p_2^\mu$. 
The constants $(f^S_\pi)_{\rm FFS}(\rho )$ and $(f_\pi^P)_{\rm
FFS}(\rho )$ are then calculated \cite{Achm 80} from the requirement
(the so--called generalised Goldberger--Treimann (GT) relation) on the
proportionality of the matrix elements of in--medium pion source operator
(~denoted by $\langle\,\widetilde{N}(\vec p_2 ) \mid \widehat{J}^a_\pi
\mid \widetilde{N}( \vec p_1)\,\rangle$~) to that of the divergence of
the purely nucleonic axial current operator $\widehat{A}^a_{\mu (N)}$,
\begin{eqnarray}
 \lim_{m_\pi^2(1+ {{\Pi^S(0,0)}\over m_\pi^2})\to 0}\!\!\!\!  -i q^\mu
 \langle \widetilde{N}(\vec{p}_2)|\widehat{A}_{\mu (N)}^a|
\widetilde{N}(\vec{p}_1) \rangle
&=& f_\pi^{\rm AP}(\rho )
 \langle \widetilde {N}(\vec{p}_2)| \widehat{J}_\pi^a |
 \widetilde {N }(\vec{p}_1) \rangle\, ,\label{fo} \\
 \widehat{A}^a_{\mu (N)}  :=
  -\bar \Psi_N \gamma_\mu\gamma_5{\tau^a\over 2}\Psi_N\quad &{\rm and}&\quad
 (\Delta_\pi)^{-1}_{\rm FFS}\,\widehat{\pi}^a =
  -\widehat{J}_\pi^a \, 
 , \label{source_pi}
\end{eqnarray}
where $\Psi_N$ stands for the nucleon field operator and 
$f_\pi^{\rm AP}(\rho )$ satisfies the relation
\begin{equation}
 \sqrt{2} f_\pi^{\rm AP}(\rho ) = 
 \lim_{m_\pi^2(1+ {{\Pi^S(0,0)}\over m_\pi^2})\to 0} 
 (\Delta_\pi)_{\rm FFS}\, 
 (-iq^\mu) \langle\, \widetilde{0}\mid 
 \widehat{A}_{\mu (\pi )}^\pm 
 \mid \widetilde{\pi}^\pm \,\rangle\, . \label{fo_rel}
\end{equation}
The latter equation is consistent with the definition of the bare pion
weak decay coupling at zero density through the condition
\begin{equation}
 \sqrt{2}f_\pi = {\lim } _{m_\pi^2 \to 0}\ 
 \left(q^2 -m_\pi^2\right)^{-1}\,(-iq^\mu)
 \,\langle\,0\mid \widehat{A}_{\mu (\pi) }^\pm 
 \mid \pi^\pm \,\rangle\, .
 \label{PCAC_cond}
\end{equation}
The limit $m_\pi^2 \to 0$ for the vacuum is the justification for
the corresponding limit in eq.\,(\ref{fo_rel}). The r.h.s.\, 
of eq.\,(\ref{fo_rel} ) is related to the axial charge --
pseudoscalar (AP) two--point function~\cite{GL 84}
\begin{eqnarray}
 {{\delta^2 Z_{\rm MF}\over 
 {\delta A_0^a(-q)\, \delta j^b(q)}}}|_{A_\mu=V_\mu=j=0,s={\cal M}} 
 & = & iq^0 
 { {f_\pi \left(1 - {{\sigma_N\rho }\over
  {f_\pi^2m_\pi^2}}\right)\delta^{ab}}\over
 {{q_0^2 - {\vec{q}\,}^2 
 { {1 -{ { D^{ii}\rho }/f_\pi^2}} \over 
 {1 + { {D^{00}\rho } /f_\pi^2}}} -{m_\pi^\ast}^2}}}
 + {\cal O}(m_\pi)\, ,
\label{two_point} 
\end{eqnarray}
via 
\begin{equation}
 f_\pi^{\rm AP}(\rho )\delta_{ab} 
 =\lim_{{\vec{q}}\to 0;{m_\pi^\ast}^2\to 0} -i q_\mu
 {{\delta^2 Z_{\rm MF}\over 
 {\delta A_\mu^a(-q)\, \delta j^b(q) }}}|_{A_\mu=V_\mu=j=0,s={\cal M}}
 = f_\pi\, \left(1 -{{\sigma_N\rho }
 \over {f_\pi^2m_\pi^2}}\right )\delta_{ab}\, .
 \label{f_walecka}                                                 
\end{equation}
Now in inserting eqs.\,(\ref{Mig}), (\ref{A_pi_vec}) and
(\ref{A_pi_0}) into eq.\,(\ref{fo_rel}) the following solution for
$(f_\pi^S)_{\rm FFS}$ (see \cite{KiRi 94} for more details) is
obtained:
\begin{equation}
 (f_\pi^S)_{\rm FFS}(\rho ) = 
 f_\pi^{\rm AP}(\rho )
 \left(1- {{\partial \Pi^S}\over {\partial q_0^2}}\right  )=
 f_\pi \,\left(1-{{\sigma_N\rho}\over {f_\pi^2m_\pi^2}}\right )
 \left(1-{{\partial \Pi^S}\over {\partial q^2_0}}\right)\, .
 \label{f_FFS}
\end{equation}
The factor $(1-{{\partial \Pi^S}\over {\partial q^2_0}})\, $ accounts
for the kinetic term of the $S$--wave pion self energy.  
Note that already the definition (\ref{A_pi_0}) implies that 
$ (f_\pi^S)_{\rm FFS}(\rho )$ has to be interpreted as
the weak axial vector decay 
constant of {\em Migdal's pion} (\ref{pi_class}). It 
can of course be constructed 
directly
from the kinetic term of the mean-field lagrangian (\ref{L_MF}), 
more precisely from the  $A_\mu^a$--$\pi^a$ interaction term 
where the axial vector source, $A_\mu^a$, should
not be mixed up with the hadronic axial current ${\widehat A}^a_{\mu (\pi)}$:
\begin{eqnarray}
 A_0^a \,i q^0 \,f_\pi\left(1 +\frac{D^{00}\rho}{f_\pi^2} 
    \right ) \pi^a
 = A_0^a\, i q^0\, \left \{ 
     f_\pi \frac{ 1 +\frac{D^{00}\rho}{f_\pi^2} }
          { 1 -\frac{\sigma_N \rho}{f_\pi^2 m_\pi^2} }\right\}
    \widetilde{\pi}^a
  =   A_0^a\, i q^0 \,(f_\pi^S)_{\rm FFS}(\rho )\, 
  \widetilde{\pi}^a \, 
     \label{A_pi_int}
\end{eqnarray}
in agreement with eq.\,(\ref{f_FFS}) --- in case eq.(\ref{Mig_dPi}) has
been inserted. If Migdal's pion field in eq.(\ref{A_pi_int}) is once
more renormalized (multiplied)
by the factor $g_\pi/g_\pi^\ast(\rho)=
(1-{ {\partial\Pi^S}\over {\partial q^2_0}})^{1/2}$, such that the 
corresponding inverse pion-propagator has weight one relative to the time-like
$\partial_0^2$ term, the physical in-medium pion decay constant 
$f_\pi^t(\rho)$ can be read from eq.(\ref{A_pi_int}) as
\begin{eqnarray}
 A_0^a \,i q^0 \,f_\pi\left(1 +\frac{D^{00}\rho}{f_\pi^2} 
    \right ) \pi^a
&=&
A^a_0\, iq^0\, f_\pi^t(\rho)\,
 \left( \frac{g_\pi}{g_\pi^\ast(\rho)}\widetilde{\pi}^a(\rho)\right)
\end{eqnarray}
in agreement with eq.(\ref{f_pi_t}).
Note that the scheme-dependence cancels out from the
combination
$(\widetilde{\pi}^a(\rho) g_\pi/g_\pi^\ast(\rho))$, such that 
the prefactor $f_\pi^t(\rho)$ 
is 
scheme-independent, too, as it should.    

The PCAC
relation within Migdal's theory is then obtained in evaluating the
matrix element of the divergence of the summed ($\widehat{A}^a_\mu =
\widehat{A}^a_{\mu (N)} + \widehat{A}^a_{\mu (\pi )}$) nucleon and
pion weak axial current between in--medium nucleon states by means of
eqs.\,(\ref{A_pi_vec}--\ref{A_pi_0NN}) in the rest frame of the pion
according to
\begin{eqnarray}
 \lefteqn{{\lim }_{\vec{q}\to 0}\ (-i q^\mu)\, 
 \langle\,\widetilde{N}(\vec{p}_2)\mid
 \widehat{A}^a_{\mu (N)} + \widehat{A}^a_{\mu (\pi )}
 \mid \widetilde{N}(\vec{p}_1)\,\rangle\nonumber}\\
 & = &  
 {\lim }_{\vec{q}\to 0}\,(-iq^\mu)
 \langle\,\widetilde{N}(\vec{p}_2)\mid \widehat{A}^a_{\mu (N)}
 \mid \widetilde{N}(\vec{p}_1 ) \,\rangle
  \qquad \qquad \qquad \nonumber \\
 && \mbox{}+ (f_\pi^S)_{\rm FFS} 
 {{q_0^2\langle\,\widetilde{N}(\vec{p}_2)\mid - \widehat{J}_\pi^a 
 \mid \widetilde{N}(\vec{p}_1)\,\rangle} 
 \over
 {\left(1-{{\partial \Pi^S}\over {\partial q_0^2}}\right)q_0^2 -
 m_\pi^2 \left(1 +{{\Pi^S(0,0)}\over m_\pi^2}\right)}}\, .
\end{eqnarray}
In inserting into the last equation the value for $(f_\pi^S)_{\rm
FFS}$ from eq.\,(\ref{f_FFS}) and in accounting for eq.\,(\ref{fo})
as well as for definition of the pion field via
eq.\,(\ref{source_pi}), one obtains the final form of the PCAC
relation in Migdal's theory as
\begin{eqnarray}
 \lefteqn{-iq^\mu
 \langle\,\widetilde{N}(\vec{p}_2)\mid
 \widehat{A}_{\mu (N)}^\pm  + \widehat{A}^\pm_{\mu (\pi )}\mid 
 \widetilde{N}(\vec{p}_1 )\,\rangle
 \nonumber}\\
 & =& \sqrt{2} f_\pi\,\left (1 -{{\sigma_N\rho}\over 
 {f_\pi^2m_\pi^2}}\right) 
  m_\pi^2 \left(1+{{\Pi^S(0,0)}\over m_\pi^2}\right)\,
 \langle\,\widetilde{N}(\vec{p}_2)\mid 
 \widehat{\pi}^\pm\mid \widetilde{N}
 (\vec{p}_1)\,\rangle
 \nonumber\\
 &=& \sqrt{2}f_\pi m_\pi^2
 \langle\,\widetilde{N}(\vec{p}_2)\mid \widehat{\pi}^\pm\mid 
 \widetilde{N}(\vec{p}_1)\,\rangle\, .
 \label{f_mat}
\end{eqnarray}
The PCAC-consistent relation (\ref{Mig_Pi}) has been used for the last
step.  Eq.\,(\ref{f_mat}) in fact means validity of PCAC as operator
relation within a finite Fermi system and thus allows the evaluation
of the GOR correlator within Migdal's approach along the line of
eq.\,(\ref{GOR_op}). As a result, also in Migdal's treatment of a
dense medium the linear decrease of the quark condensate with density
is recovered,
\begin{equation}
 -2m_q\langle\,\widetilde{0}\mid \bar q q\mid \widetilde{0}\,\rangle
 = \, f_\pi^2 m_\pi^2 \, 
 {1\over {1+ { {\Pi^S(0,0)}\over {m_\pi^2}} } }
 = f_\pi^2m_\pi^2 \,\left(1-{{\sigma_N\rho}\over
  {f_\pi^2m_\pi^2}}\right)
 = \left(f_\pi^t(\rho)\right)^2 \left (m_\pi^\ast\right)^2\, .
\end{equation}
Thus again the factor $1/1+{{\Pi^S(0,0) }\over {m_\pi^2}}$ is
converted back to the rescaling factor of the pion field at zero
matter density $\pi^a$ to the Migdal field $\widetilde{\pi}^a$ at
finite matter density from eq.\,(\ref{pi_class}) --- in agreement with
the generating functional formalism~\cite{ThWi 95}. In this way the
QCD predicted linear decrease of the quark condensate in the medium
is, as expected, recovered.  

\section{\bf Summary  and discussion}

To summarize we wish to stress that it is the quark
condensate that depends linearly on density rather than the $S$--wave
pion polarization function $\Pi^S(q_0, \vec q = 0 )$ as currently used
in the literature.  The quantity $\Pi^S(q_0, \vec q = 0 )$ in fact
contains higher powers of $\rho $ resulting from the expansions
(\ref{Mig_dPi}, \ref{Mig_Pi})
\begin{equation}
 {{\Pi^S(0,0 )}\over m_\pi^2} = {1\over 
 {1-{{\sigma_N\rho  }\over {f_\pi^2m_\pi^2}} } } -1\, =\,
 {{\sigma_N\rho }\over {f_\pi^2m_\pi^2}}
 \left(1 + {{\sigma_N\rho }\over {f_\pi^2m_\pi^2}}
 + \left({{\sigma_N\rho }\over {f_\pi^2 m_\pi^2}}\right)^2
 + \cdots \right)\, \approx -{{\partial \Pi^S}\over {\partial q_0}}|_{q_0^2=0}
\, .
\end{equation}
\noindent

In an early work ref.\,\cite{Achm 80} the quantity $f_\pi^{\rm
AP}(\rho) $ in eq. (\ref{fo_rel}) was approximated by the bare pion
weak decay constant $f_\pi$. This approximation implies the absence of
the factor $1\mbox{$-$}{ {\sigma_N\rho }\over {f_\pi^2m_\pi^2}}$ in
the expression for $(f_\pi^S)_{\rm FFS}(\rho) $ in
eq.\,(\ref{f_FFS}) and leads therefore to the following changes in
the PCAC condition of eq.\,(\ref{f_mat}):
\begin{eqnarray}
 \lefteqn{f_\pi m_\pi^2 \left(1-{{\sigma_N\rho}\over  
 {f_\pi^2m_\pi^2}}\right)
 \left(1+{{\Pi^S(0,0)}\over m_\pi^2}\right)
  = f_\pi m_\pi^2 \nonumber}\\
 &&\Longrightarrow 
 f_\pi m_\pi^2 \left(1+{{\Pi^S(0,0)}\over m_\pi^2}\right) 
 =  f_\pi m_\pi^2 
 { 1\over {1-{{\sigma_N\rho}\over {f_\pi^2m_\pi^2}} }}\, .
\end{eqnarray}
With that, the quark condensate resulting from the evaluation of the
GOR correlator (\ref{GOR_op}) in that approximation would be
proportional to
\begin{equation}
 -2m_q \langle\,\widetilde{0}\mid \bar q q\mid \widetilde{0}\,
 \rangle\ {''\!\!=''}\ f_\pi^2m_\pi^2 \left(1+ {{\Pi^S(0,0)}
 \over m_\pi^2}\right ) = 
 {{f_\pi^2m_\pi^2}}/\left( {1 - {{\sigma_N\rho}\over 
 {f_\pi^2m_\pi^2}}}\right )\, ,
\end{equation}
and thus increasing instead of decreasing with density.  This
observation underlines once more the necessity for a careful and
consistent treatment of PCAC before exploiting it for the calculation
of measurable and thus model-- or scheme--independent QCD quantities.
There is no real need to refer to PCAC on the composite hadron level
to calculate quantities on the current quark level.  However, if one
prefers to work in this scheme, one has to apply it without
violating consistency relations having their roots in the current
quark dynamics. 
As presented above, the big virtue of the generating
functional scheme is that it allows for a transcription of the source
structures from the current quark level of the underlying QCD to the
effective hadron level without invoking any model assumptions of how
the composite hadrons are built up from the quarks.

\section{\bf Acknowledgement}
This work was partly supported by the Deutsche Forschungsgemeinschaft (SFB 201).

\end{document}